\documentclass[showpacs,preprint]{revtex4}
\begin{document}

\title{ Magnetic hysteresis loop as a probe to distinguish single layer from many layer graphitic structure.}
\author{K. Bagani, B. Ghosh, M. K. Ray, N. Gayathri$^1$, M. Sardar$^2$ and S. Banerjee\footnote{email:sangam.banerjee@saha.ac.in}}
\address{Surface Physics Division, Saha Institute of Nuclear Physics, 1/AF Bidhannagar, Kolkata-700064, India\\
$^1$Material Science Section, Variable Energy Cyclotron Centre, 1/AF Bidhannagar, Kolkata-700064, India\\ 
$^2$Material Science Division, Indira Gandhi Center for Atomic Research, Kalpakkam 603 102, India}

\begin{abstract} In this report we have pointed out that magnetic hysteresis loop can be used as a probe to distinguish a single layer from a many layer graphitic structure. Chemically we have synthesized graphitic oxide (GO) and reduced graphitic oxide (RGO) for this investigation. We observe ferromagnetic like hysteresis loops for both GO and RGO below a certain applied critical magnetic field and above this critical field we observe cross-over of the positive magnetization to negative magnetization leading to diamagnetic behaviour. This cross-over is more dominant for the case of many layer graphitic structure. Upon annealing of GO in air the critical cross-over field decreases and the magnetization increases for multilayer graphitic structure. Possible reasons for all these observations and phenomena is presented here.

\end{abstract}
\maketitle

\section{INTRODUCTION:}
Low-dimensional graphitic structure such as graphene have attracted much research interest because of
their unique electronic \cite{Geim,banerjee}, mechanical \cite{mechanical,mechanical2}, thermal \cite{thermal} and magnetic \cite{sepioni,magnetic2,magnetic3,Esquinazi} properties to name few. Here in this paper we would like to investigate the magnetic behavior of graphitic oxide (GO) and reduced graphitic oxide (RGO) when it is evolving from single layer or very few layers to many layers tending towards bulk form. It is well known that graphite in a bulk form (3d-structure) exhibits diamagnetic behavior. But, it has been observed that when highly oriented pyrolytic graphite (HOPG) is irradiated with proton beam it exhibits ferromagnetic behaviour \cite{Esquinazi}. The observation of ferromagnetism in HOPG and its related materials such as graphene \cite{Wang} is surprising because we know that in general the magnetic ordering is typically observed in materials with partially filled d or f shell electrons. Magnetism in zero-dimensional graphene nanofragments, one-dimensional graphene nanoribbons and defect-induced magnetism in graphene and graphite have been reviewed recently by Yazyev \cite{Yazyev}. Many theoretical works already exists to explain unusual magnetic property in graphene related materials and are cited in our recent work \cite{cond-mat}. In this report we would like to show that magnetic hysteresis as a function of field and temperature can be used as a probe to distinguish single layer graphitic structure from many layer graphitic structure tending towards bulk limit.

\section{EXPERIMENTAL DETAILS:}
Graphitic Oxide (GO) was synthesized using modified Hummers method \cite{Hummers}. To prepare GO we used graphite powder, conc. H$_2$SO$_4$, NaNO$_3$ and KMnO$_4$. The reaction of H$_2$SO$_4$ with strong oxidizing agents such as NaNO$_3$ and KMnO$_4$ results in increasing the interlayer separation of the carbon layers due to incorporation/intercalation of oxygen atoms. This reaction leads to formation of graphite oxide flakes. The solution was further ultrasonicated and centrifuged. Six different samples were collected at various stages: samples were filtered from the solution before reducing and we labelled it as sample A, after reducing with hydrazine hydrate as sample B, after annealing at 150 $^o$C and 450 $^o$C as sample C and sample D respectively and finally samples collected from supernatant without reduction as sample E and upon reduction of the supernatant particles with hydrazine hydrate as sample F. The supernatant and the filtered samples were washed with water and ethanol and was further collected and dried in vacuum drying oven at 80$^{0}$C to obtain dry powder samples. These samples were characterised by UV-VIS, FTIR, Raman spectroscopy, and SEM as described in \cite{cond-mat}. 

Magnetic measurements were carried out using SQUID MPMS XL (Quantum Design). Magnetization data as a function of temperature and magnetic hysteresis data at three different temperatures 10 K, 100 K and 300 K were taken. Thermal gravimetric analysis (TGA) for qualitative estimation of oxygen content in the samples were also carried out.

\section{RESULTS:}
In fig. 1 we show magnetization versus temperature for sample A, B, C and D and the appropriate samples are labelled in the figure. We show zero field cooled (ZFC) and field cooled (FC) measurement at 100 Oe for all the samples while warming. We see that for all the samples there exist clear bifurcation in the ZFC-FC curve around room temperature (300K). The amount of bifurcation increases upon reduction with hydrazine (sample B) or upon annealing at high temperature (sample D). We also see that magnetization value (moments) increases upon annealing and upon chemical reduction of graphitic oxide (GO). We observe that upon annealing GO at 450 $^o$C (sample D) the moment increases more than the chemically reduced sample (sample B). These are important observation and we would like to discuss these in discussion section below. In fig. 2 (a, b and c) we show the magnetization versus the applied field taken at 10 K, 100 K and 300 K for GO and RGO samples and we observe a distinctly sharp cross-over of magnetization from positive to negative value and vice versa in the negative direction of the applied field beyond a certain critical field for as prepared GO and annealed GO samples (sample A, C and D) in 100 K and 300 K data.
The saturation moment of the as prepared GO (sample A) at $T=100$ and $300$K are $1.75 \times 10^{-2}$ emu/gm and $1.2 \times 10^{-2}$ emu/gm respectively. The crossover field, i.e., the field at which
magnetization crosses over to negative values,  is 3 Tesla and 1.2 Tesla at temperatures of 100 and 300 K.
Beyond the crossover field, the magnetization is linear in field, with a diamagnetic susceptibility of
$\chi_{dia}= 8 \times 10^{-7}$ emu/gm/Oe and $1.325 \times 10^{-6}$ emu/gm/Oe for temperature 100 K and 300 K respectively. This amounts to a 40 \% increase
in the diamagnetic susceptibility going from T=100 K to 300K. Diamagnetic suceptibilty is expected to come from the $\pi$ electrons/carriers from the relatively undistorted planar regions (sp$^2$ bondend) free from functional groups like O or OH. The increase in diamagetic susceptibilty with increase in temperature indicates a gap in the electronic spectrum i.e., GO is electrically insulating
with a small gap.  The magnetization data of sample B (got after chemically reducing sample A, as described in the text) is interesting. We notice that the magnetization do not saturate even at H=7 Tesla at 100 K. The magnetic moment at H=7 Tesla is 0.105 emu/gm. The ratio of positive saturation magnetization at T=100 K of sample B and sample A is about 6. This is large increase in the magnetization of graphene oxide upon chemical reduction. At T=300 K, the crossover field of sample B is $\sim$ 5 Tesla.  The corresponding value of crossover field of sample A (at T=300 K) is 1.2 Tesla. So the crossover field at 100 K for sample B must be higher than the maximum field used in the experiment (7 Tesla). It should be also noticed that crossover field is higher in sample B (5 Tesla) compared to sample A (1.2 Tesla).
The diamagnetic susceptibility at high field of sample B at T=300 K, is $1.285 \times 10^{-6}$ emu/gm/Oe. This means a 50\% increase in diamagnetic susceptibilty of GO after chemical reduction. The chemical reduction process gets rid of O and OH 
groups from graphene surfaces leading to release of many $\pi$ electron carriers which were bonded to the radicals. Reduction process also leads to restoration of planar sp$^2$ bonded regions. Both of these probably causes substantial increase in $\pi$ electron diamagnetism.
Since diamagnetism of RGO is more than GO, the higher crosover field of RGO compared to GO can come only because of
much higher para/ferro-magnetic moments in RGO compared to GO ( a factor of 6 ). The relevant question here is how the reduction process of GO can lead to such increase of moment? Now let us compare this observations with that of sample C and D (which were obtained by heating sample A at 150 $^o$C and 450 $^o$C). The diamagnetic susceptibility ($\chi_{dia}$) and saturation magnetization ($M_{sat}$) has the following order,
$\chi_{dia}(A) \approx  \chi_{dia}(C) < \chi_{dia}(D) $ and  $M_{sat}(A) \approx  M_{sat}(C) < M_{sat}(D)$ .
The crossover field progressively decrease from A to C to D. Summarizing : (1) With increase in annealing temperature (going from A to C and D) there is an increase of para/ferro-magnetic moment. (2) Along with this there is also an increase in diamagnetic susceptibility. The increase in diamagnetic susceptibility after annealing can be
roughly understood as due to removal of many O and OH groups from the surface and consequest recovery/increase in area
of sp$^2$ bonded planar zones. Here again the main puzzle is the reason for increase in net para/ferro-magnetic moment after annealing.
So both chemical reduction done at low temperatures and annealing at high temperatures increase net magnetic moment.
We also observe the saturation magnetic moment (the value of magnetization at H=7 Tesla) and the crossover field for the reduced sample (B)
is higher compared to sample C and D. In other words chemical reduction is more effective compared to annealing in generating extra magnetic moment on graphene oxide. In fig. 3(a,b and c) we show hysteresis loops magnified around origin for GO (sample A) and RGO (sample B) taken at 10 K, 100 K and 300 K respectively. We could clearly observe that upon reduction of GO the moment increases. The coercive field decreases with increase in temperature for both the samples A and B. The reduction of coercivity with increase in temperature is an important result as this is oppose to what we have reported recently for single layer graphene structure. This important aspect will also be discussed in the section below. In fig. 4(a, b and c) we show for further clarity that RGO have more moments than GO for sample collected by filtering (fig. 4(a)). The cross-over in the case of GO happens at lower temperature and lower fields fig. 4(b and c) than RGO.

In fig. 5 we show magnetic hysteresis curves for GO and RGO (sample E and F) collected and dried from the supernatant liquid. Since the sample is collected from the graphitic suspension, one can assume that the particle sizes are smaller and contains much less number of layers tending to few layers or even tending to single layer (graphene).  We can clearly see from the hysteresis curves for this supernatant samples the coercivity on the other hand increases upon increase in temperature contrary to all the above samples (samples A to sample D). This we have reported recently in another set of sample \cite{AIP}. This behaviour (sample E and F) that we have found is  reproducible. We have also observed that in the case of supernatant samples the moments of GO sample is more than chemically reduced samples unlike the above samples A to D. In fig. 6 we show TGA curves for sample A, B and D. We see that the sample D have the least oxygen content i.e., observation of least weight loss as we increase the temperature. This indicates that the sample D is the most reduced sample.

\section{DISCUSSION:}

Let us point out the main features observed in the result section above and they are as follows:\\
(We seperate the set of samples as two types ie., (1) sample filtered from the solution (type I) and (2) Samples collected from supernatant (type II)).\\
We observe:\\
(1) For the case of type I the magnetization/moment of as-such collected GO samples after filtering are {\bf\textquotedblleft less"} than chemically reduced or annealed samples.\\
(2) For the case of type II the magnetization/moment of as-such collected GO samples from the supernatant are {\bf\textquotedblleft higher"} than chemically reduced  samples.\\
(3) For the case of type I samples the coercivity of the magnetic hysteresis curve {\bf\textquotedblleft decreases"} upon increasing the temperature. This is the expected behaviour of sparsely distributed (weak interaction) single domain magnetic moments \\
(4) For the case of type II samples the coercivity of the magnetic hysteresis curve {\bf\textquotedblleft increases"} upon increasing the temperature. This is highly unusual, but as pointed out earlier is a very reproducible result \cite{cond-mat}.\\

These opposing behaviour observed in these two type of sample is being reported for the first time to our knowledge. For our convenience we shall discuss the type II sample first and then the type I samples 

Recently we have proposed that for the case of single layer graphitic structure such as graphene oxide and graphene (type II structure) the magnetic property such as coercivity of hysteresis curve is governed by the inherent presence of ripple/wrinkles in these single layer sheets. It has been observed that these ripples/wrinkles diminishes upon heating and this diminishing of ripple/wrinkles as a function of increasing temperature was invoked in reference \cite{cond-mat} to explain the increase in coercivity of magnetization as a function of temperature. For the type II samples GO has more
paramagnetic moment compared to RGO \cite{AIP}. The origin of moments could be \cite{cond-mat} due to the presence of OH groups and other defects such as uncompensated sites, vacancies etc.,. It is  natural to assume that the reduction process removes many such small moment carrying OH clusters from GO. Upon reduction many OH groups are removed and hence we can consider less clustering of OH groups and isolated OH groups are scattered all over the sheet. Since GO has more OH groups than RGO, the net magnetic moment of GO is higher than RGO.  An OH can bind to C atom on any sublattice. If there is an local imbalance in the number of OH group attched to C atoms on sublattice A and sublattice B then there is a local moment formation by Lieb's theorem \cite{Lieb}. We have to remember that the local spin correlation between nearest neighbour C atoms is antiferromagnetic in nature (superexchange) due to large unscreened coulomb repulsion between the $\pi$ electrons. So a local level cluster of OH groups on graphene surface will lead to some uncompensated but large moment around such a region. This was proposed by us to be the reason for magnetism in GO (sample E).
In the reduction process many such OH groups will be removed leading to removal of many moments around them. This process also increases the diamagnetism of the materials because of the excess diamagnetic susceptibilty of the $\pi$ electron, which were earlier bonded to an OH group. This was the reason for reduction of paramagnetic moment in sample F. The magnetism of all samples is composed of three parts: (a) Backgroud diamagnetism of the $\pi$ electrons, (b) Paramagnetism at all temperatures coming from small moments due to vacancy, dangling bonds or very small OH clusters on graphene surface and (c) There are also some resonably large single domain magnetic moments, presumably coming from either exposed zigzag edges or large patches of OH groups on the surface. It is (c) which are responsible and important for magnetic irreversibilty and hysteresis. In the reduction process many OH groups are removed and hence net magnetic moment decreases.
At low temperatures these isolated moments randomly
placed in an insulating matrix with very little interaction between them due to highly crumpled structure behaves like paramagnets. The point number (2) and (4) mentioned above in the begining of this section has been explained in detail in reference \cite{cond-mat} when the graphitic structure is like graphene oxide or graphene. The curious increase of coercivity with increase in temperature
was argued by us as due to reduced average crumpling on single or few layer graphene oxide at higher temperature. Less crumpling leads to higher locally averaged magnetic anisotropy energy of such single domain large moments. Since coercivity of such materials is proportional to the average anisotropy energy of the single domain moments, the coercivity increases with increase in temperatures. We also pointed out the insulating nature of such materials with small gap ensuring an increase in number of carriers with increase in temperature. This liberated carriers can also bring in long range ferromagnetic correlation and coercivity can increase because of that.

Now what happens to magnetization when we do not have single or few layer graphitic powder but many layered graphitic powder tending towards the bulk graphitic structure. First let us address the decrease in the coercivity in the hysteresis curve upon increasing the temperature. This is an usual behaviour expected in Stoner - Wolfarth model \cite{Stoner}, which applies to a collection of single domain non-interacting magnetic moments.

We have recently shown that when the graphene is loosely held on HOPG, then there exist ripple and upon strong adherence of graphene with the bulk HOPG substrate the ripple vanishes \cite{JPCripple}. Thus multilayer graphitic structure will not have any ripple formation like one generally observe in a single layer graphitic structure. Thus the many layer graphitic structure will behave more like sparsely distributed nanomagnetic system. That is why the coercivity has the usual decrease with increase in temperatures.

Observation of increase in magnetization value upon chemical reduction or annealing at high temperature in air for the case of multilayer graphitic structure (sample B, C and D) is very interesting. Recently many effort has been made theoretically and experimentally to identify the structure of GO and it is now believed that the oxygen and the hydroxyl group are bonded to graphene in a definite fashion such that upon reduction or annealing (ie., removal of oxygen or hydroxyl) leads to unzipping of the graphene structure leading to fragmentation of the graphene layers leaving many zigzag edges being exposed. It appears from our experimental observation that unzipping is more prominent for the case of multilayered graphitic structure than single layered structure and this is mainly due to development of strain upon reduction/unzipping in multilayered structure and the strain for the case of single layer is less than multilayer which may be obvious due to the nature of the structure (i.e., for single layer the strain developed across the sample is less than in multilayered samples). These development of strain and unzipping of the layers leads to heavy fragmention of the bulk graphitic structure to large smaller fragments exposing large number of zig-zag edges. If we invoke the occurrence of magnetization due to zig-zag edges \cite{zig-zag-mag}, then we can explain the increase of magnetization upon chemical reduction or annealing in air. We are working on this aspect of unzipping and shall be reporting elsewhere but however many prilimainary works have been reported by other \cite{unzipping, unzipping1, unzipping2}.
At present our discussion can only be speculative, but more theory and experiments are looking for ordered local arrangement of O and OH groups on graphene.  There could be local low energy structures, where OH group is attached to a chain of C atoms. The OH groups are above and below the graphene surface  over carbon atoms at A and B sublattices.
For single layer GO, all such OH groups  on both sides will be randomly removed, and  it is unlikely to create a local imbalance in A and B sublattice sites to which OH is attached. For single or few layer graphene oxide, when reduced there is only minute increase/decrease of moment. But  a multilayer graphene when reduced may have uneuqal reduction on different sides of graphene surface. This will create a large extra moment.

Another interesting observation by us is that the low field magnetization of sample D is more than sample B, whereas 
high field magnetization of B is higher than sample D.
Since upon annealing at 450 $^o$C, the graphitic flakes disintegrates into smaller sizes and hence the size of the  graphitic flakes in sample D are much smaller that in sample A or sample B. This leads to many small segment of edges carrying moments. Hence, the low field magnetization is higher in sample D than sample B. Sample B has many
large sized moments also, but they are blocked at low field. At high field these large moments align along the external field giving higher magnetization in sample B (chemically reduced sample). In other words, the moments at the edges align at low fields (sample D) and the moments due to patches in chemically reduced samples (sample B) align at higher fields.

Finally we point out again that the crossover of magnetization towards negative values happens over a very small field interval.
This almost discontinuous change of magnetization, looks like some kind of magnetic transition. More work is needed to
understand this phenomena.

\section{CONCLUSION:}
In conclusion, we have shown for the first time that for "single or few layer" graphitic oxide (GO) and reduced graphitic oxide (RGO) the magnetization is larger for GO than RGO and for both these samples the magnetic coercivity increases with increase in temperature. Whereas, for multilayer GO or RGO the above two properties are opposite i.e., RGO have larger mangetization than GO and the magnitude of magnetic coercivity decrease with increase in temperature.

\newpage
\centerline{\bf Figure Captions}

\vspace{0.5 cm}

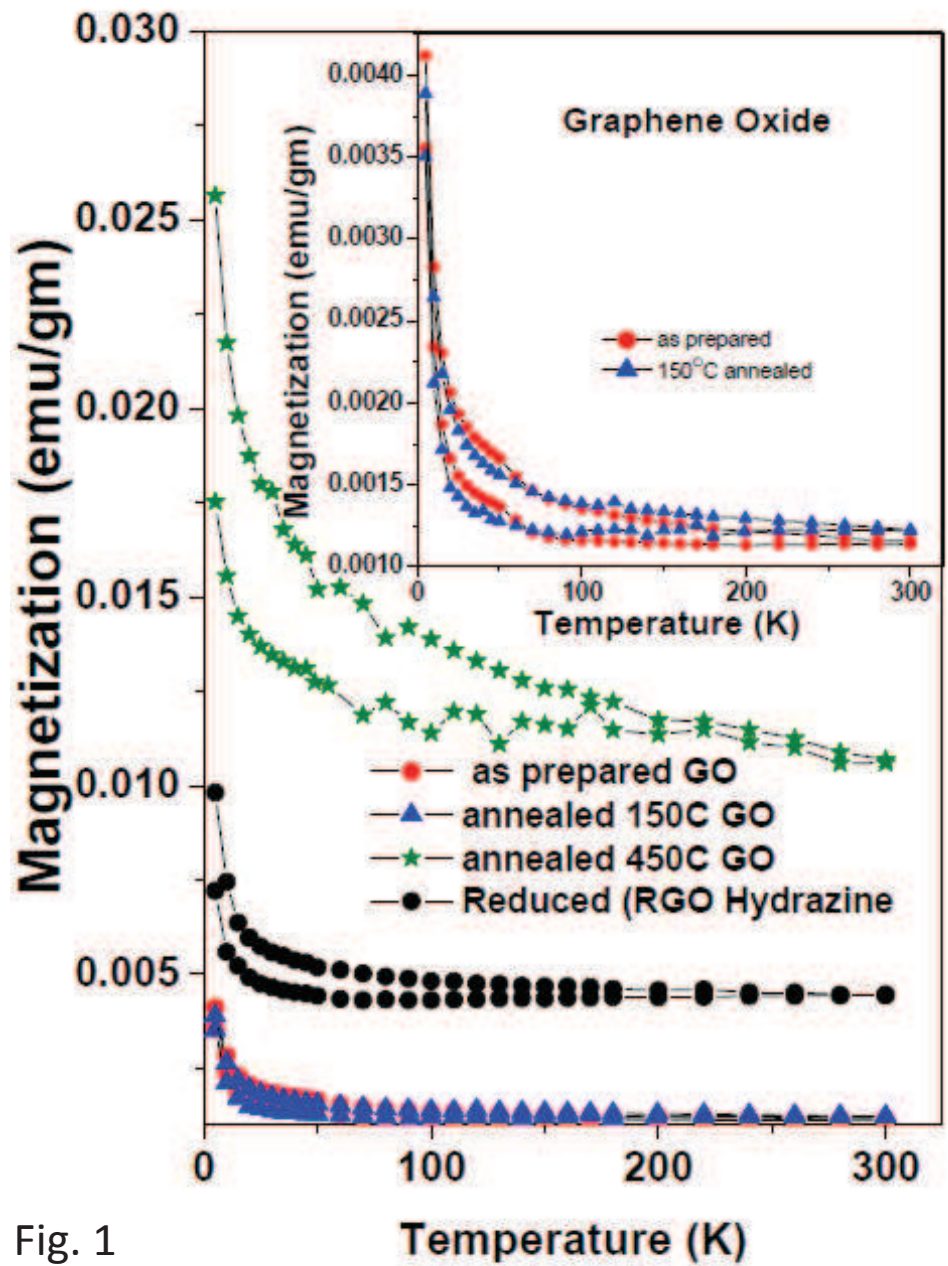
{\bf Fig. 1} ZFC and FC magnetization versus temperature for sample A, B, C and D and the appropriate samples are labelled in the figure.\\

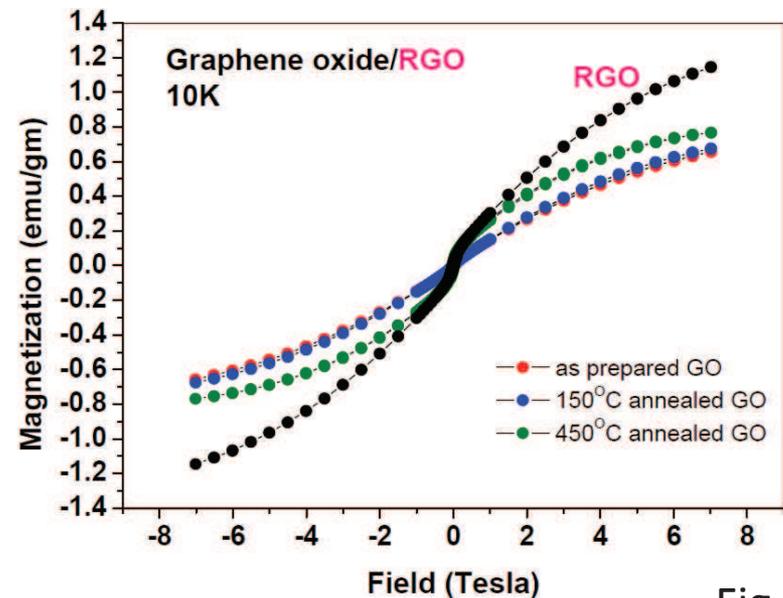
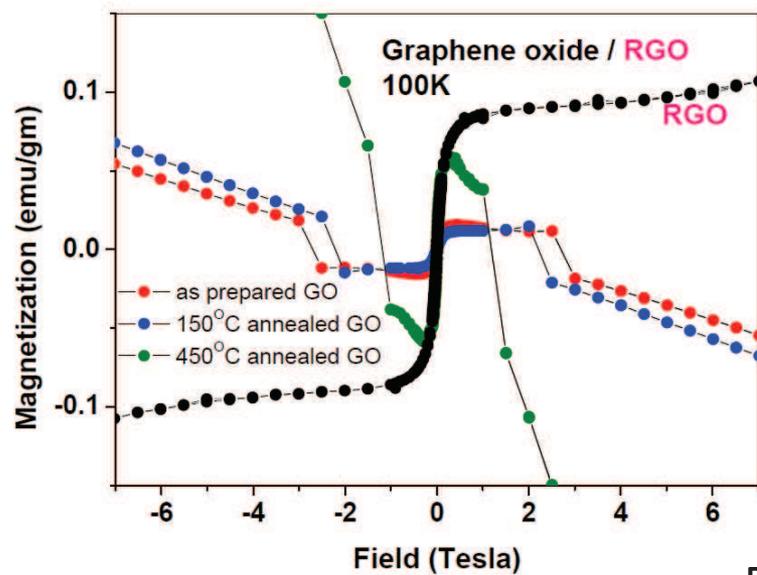
{\bf Fig. 2} Magnetization versus the applied field taken at (a) 10 K, (b) 100 K and (c) 300 K for GO and RGO samples and we observe a distinctly sharp cross-over of magnetization from positive to negative value and vice versa in the negative direction.\\

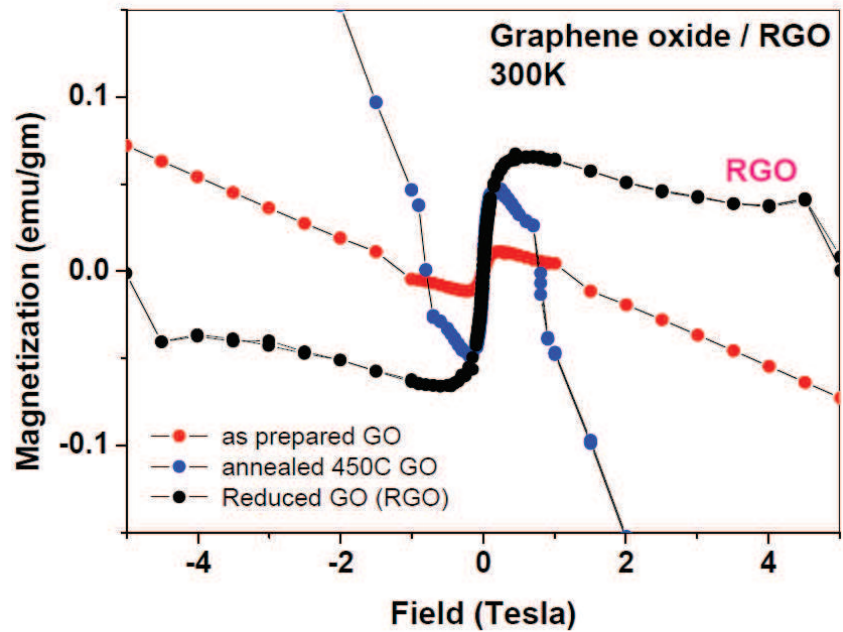
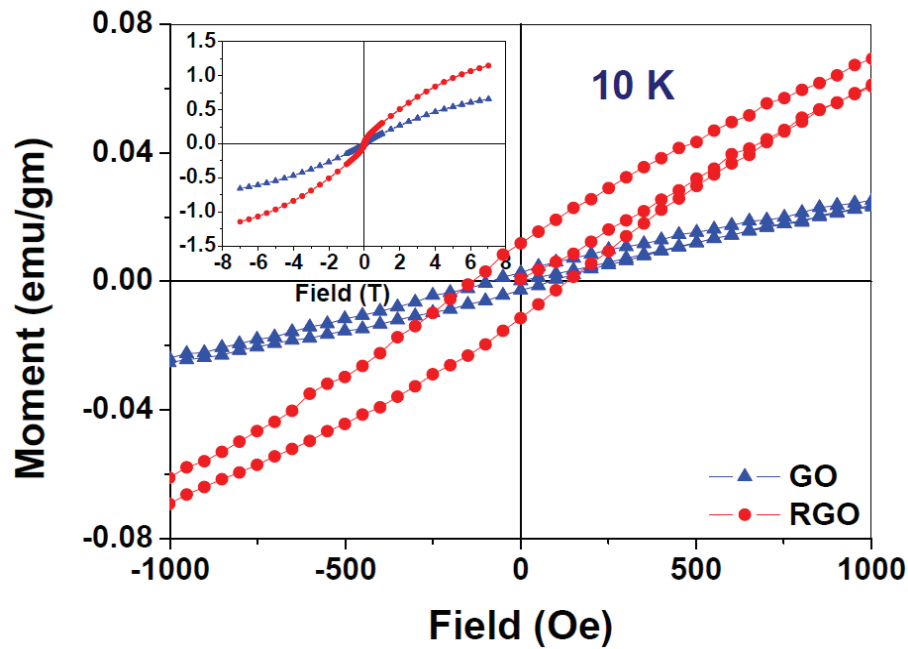
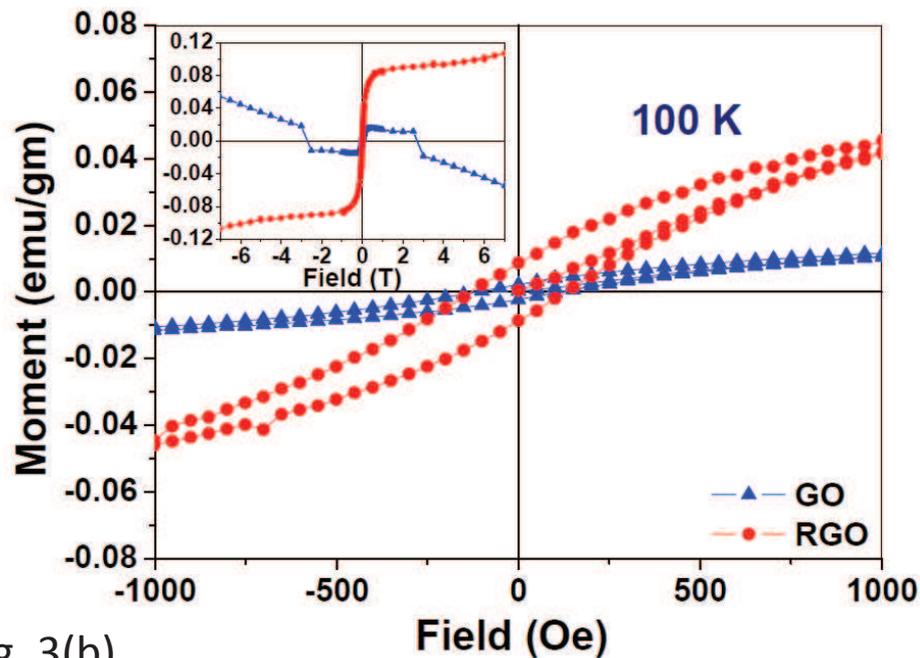
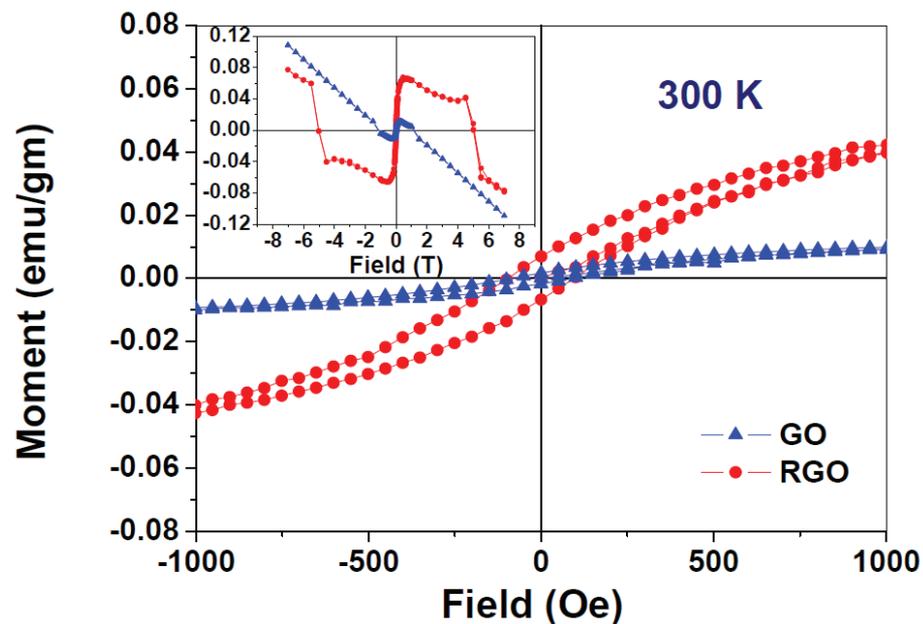
{\bf Fig. 3} Hysteresis loops magnified around origin for GO (sample A) and RGO (sample B) taken at (a) 10 K, (b) 100 K and (c) 300 K respectively.\\

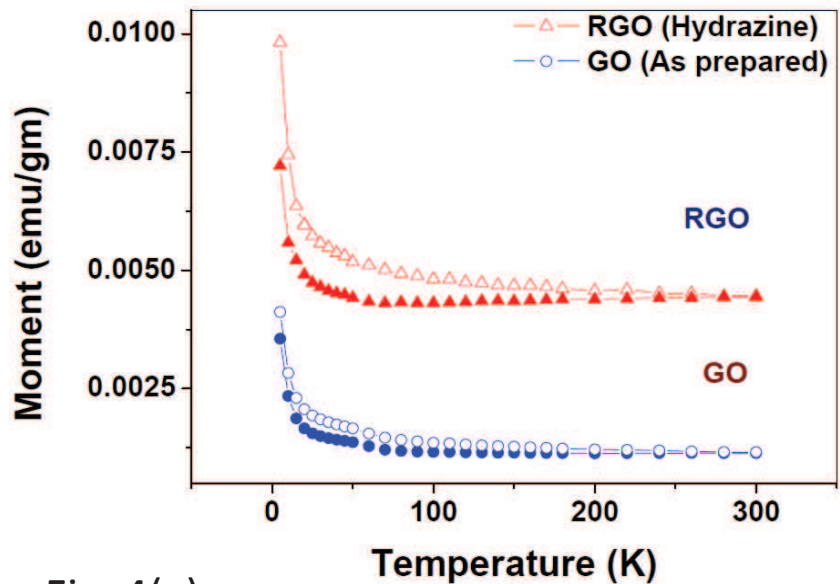
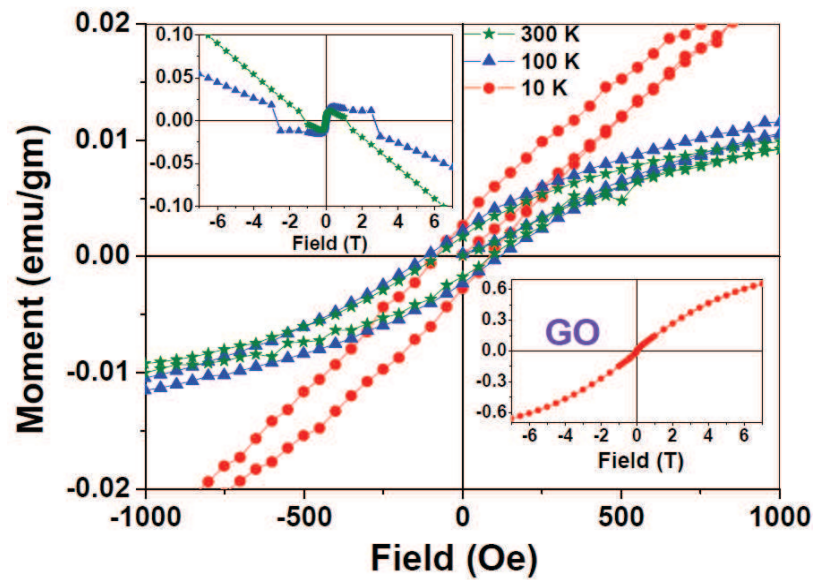
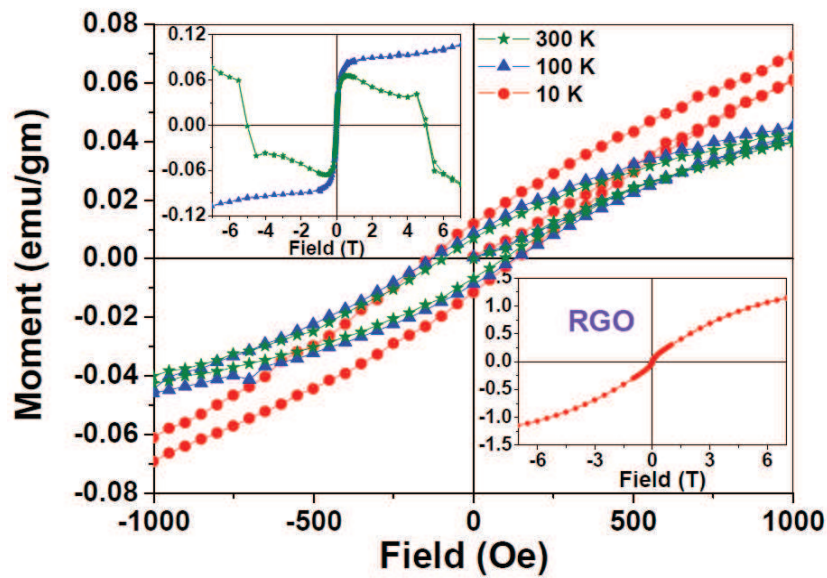
{\bf Fig. 4} For further clarity to show that RGO have more moments than GO for sample collected by filtering.\\

{\bf Fig. 5} Magnetic hysteresis curves for GO and RGO (sample E and F) collected and dried from the supernatant liquid.\\

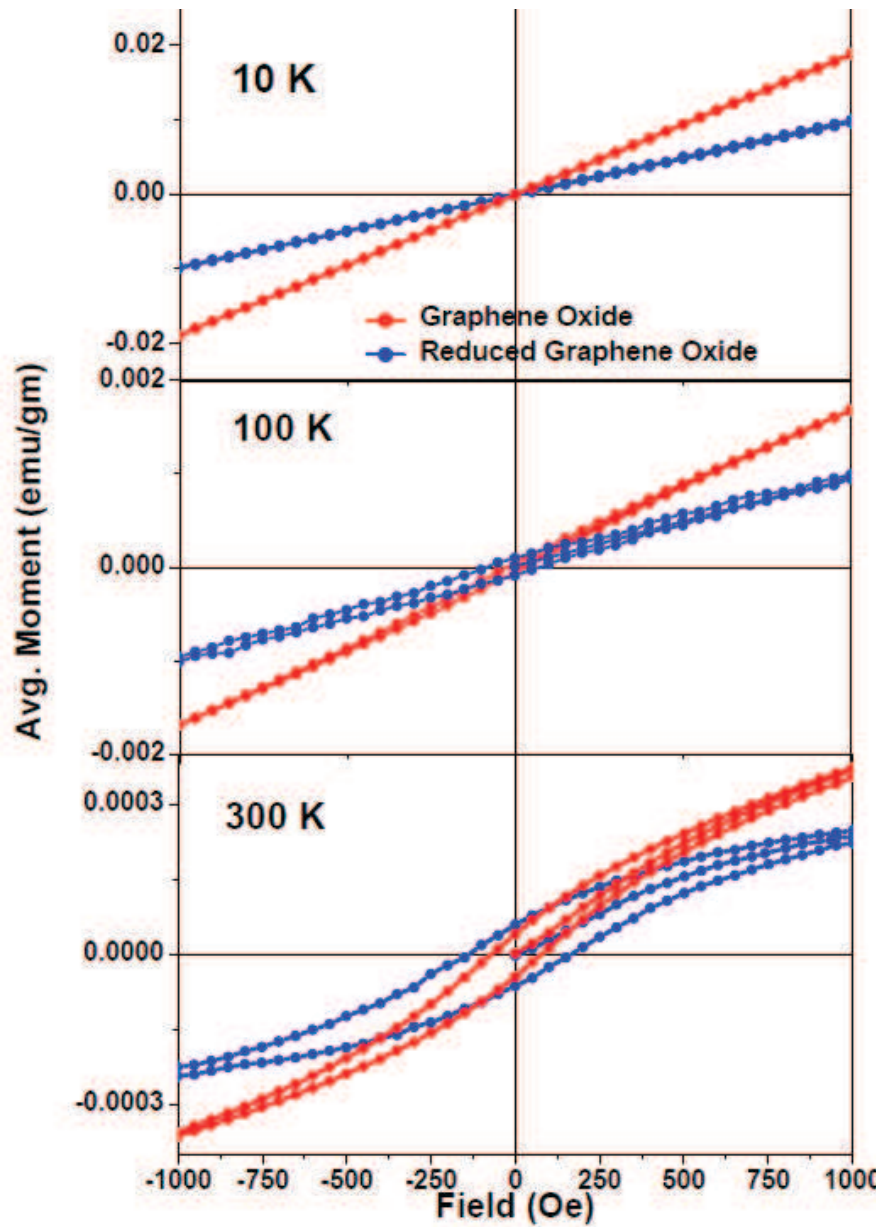
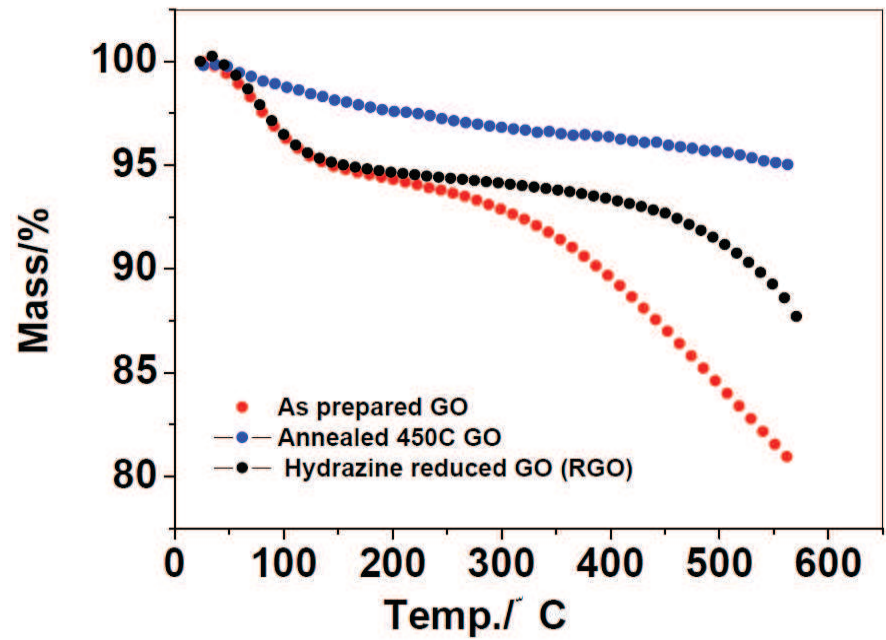
{\bf Fig. 6} TGA curves for sample A, B and D

\vspace{0.5 cm}


\begin{thebibliography}{99}

\bibitem{Geim} H. C. Neto, F. Guinea, N. M. R. Peres, K. S. Novoselov and A. K. Geim; Rev. Mod. Phys. {\bf 81}, 109  (2009).  A.K.Geim and K. S. Novoselov; Nature Mater. {\bf 6}, 183 (2007).  A. K. Geim and A. H. MacDonald; Phys. Today {\bf 608}, 35(2007).

\bibitem{banerjee}S. Banerjee, M. Sardar, N. Gayathri, A. K. Tyagi, and Baldev Raj, Phys. Rev. B {\bf 72}, 075418 (2005); S. Banerjee, M. Sardar, N. Gayathri, A. K. Tyagi, and Baldev Raj, Appl. Phys. Lett.88, 062111 (2006)
\bibitem{mechanical} J. W. Suk, R. D. Piner, J. An, and R. S. Ruoff, ACS Nano, {\bf 4}(11), 6557–6564 (2010)
\bibitem{mechanical2} I. W. Fran, D. M. Tanenbaum, A. M. vanderZande and P. L. McEuen,  J. Vac. Sci. Technol. B  {\bf 25}(6), 2558(2007): A. R. Ranjbartoreh, B. Wang, X. Shen and G. Wang, J. Appl. Phys. {\bf 109}, 014306(2011)
\bibitem{thermal}  A. A. Balandin, S. Ghosh, Wenzhong Bao, I. Calizo, D. Teweldebrhan, F. Miao and C. N. Lau, Nano Lett., {\bf 8}, 3(2008); K. Saito, J. Nakamura and A. Natori, Phys. ReV.B, {\bf 76}, 115409(2007).

\bibitem{sepioni} M. Sepioni, R. R. Nair, S. Rablen, J. Narayanan, F. Tuna, R. Winpenny, A. K. Geim  and I. V. Grigorieva; Phys. Rev. Lett. {\bf 105} 207205 (2010).

\bibitem{magnetic2} D. Jianga, B. G. Sumpter and S. Dai, J. Chem. Phys., {\bf 127}, 124703(2007).
\bibitem{magnetic3} H. S. S. Ramakrishna Matte, K. S. Subrahmanyam, and  C. N. R Rao,J. Phys. Chem. C {\bf 113} 9982 (2009).
\bibitem{Esquinazi} P.Esquinazi, D. Spemann, R. Ho ¨hne, A. Setzer, K. H. Han, T.  Butz; Phys. ReV. Lett. {\bf 91}, 227201(2003).

\bibitem{Wang} Y. Wang, Y. Huang, Y. Song, X. Zhang, Y. Ma, J. Liang and Y. Chen; Nano Letters  {\bf 9}, 1, 220-224(2009).

\bibitem{Yazyev} O.V. Yazyev  and L. Helm;  Phys. Rev. B {\bf 75} 125408 (2007), O.V. Yazyev, Rep. Prog. Phys. {\bf 73} 056501 (2010)

\bibitem{cond-mat} K. Bagani, A. Bhattacharya, J. Kaur, A. Rai Chowdhury, B. Ghosh, M. Sardar and S. Banerjee, arXiv:1302.3336v1 [cond-mat.mtrl-sci] 14 Feb 2013

\bibitem{AIP} J. Kaur, A. Rai Chowdhury A. Bhattacharya, B. Ghosh, M. Sardar and S. Banerjee, AIP Conf. Proc. {\bf 1447}, 1233 (2012); doi: 10.1063/1.4710457

\bibitem{Hummers} W. Hummers, R. Offeman; J. Am. Chem. Soc. 1958, 80, 1339

\bibitem{Lieb} E. H. Lieb, Phys. Rev. Lett. 62, 1201 (1989).

\bibitem{Stoner} E. C. Stoner and E. P. Wohlfarth,  Phil. Trans. R. Soc. Lond. A {\bf 240} 599 (1948)

\bibitem{JPCripple} S. Panigrahi, A. Bhattacharya, S. Banerjee and D. Bhattacharyya, J.Phys.Chem.C  {\bf 116}, 4374-4379(2012).


\bibitem{zig-zag-mag} H. Kumazaki and  D.  S. Hirashima; J. Phys. Soc. Jpn., {\bf 77}(4), 044705(2008), S. Bhowmick and V.  B. Shenoy, J. Chem. Phys., {\bf 128}, 244717(2008), J. Jung and A. H. MacDonald; Phys.  Rev. B {\bf 79}, 235433  (2009).


\bibitem{unzipping} D. Pan, J. Zhang, Z. Li, and M. Wu, Adv. Mater. {\bf 22}, 734–738(2010)

\bibitem{unzipping1} J. Li, K. N. Kudin, M. J. McAllister, R. K. Prud’homme, I. A. Aksay and R. Car, Phys. Rev. Lett, {\bf 96}, 176101(2006)

\bibitem{unzipping2} T. Sun and S. Fabris, Nano Lett., {\bf 12}, 17−21(2012).


\end{thebibliography}
\end{document}